\documentclass[12pt]{article}
\usepackage[cp1251]{inputenc}
\usepackage[russian]{babel}
\begin{document}
\title{ On quantization of systems with second class constraints}
\author{A.G. Nuramatov and  L.V. Prokhorov}
\date{}
\maketitle{}
\renewcommand {\refname}{References}
\renewcommand {\abstractname}{Abstract}
\begin{abstract}
 It is shown that quantization of the dynamical systems with
 second class constraints actually can be reduced to
 quantization of the systems with first class constraints. The motion
 of the non-relativistic particle along the plane curve and
 on a surface is considered. The results coincide with those of the "thin layer
 method". Influence of the non-physical variables on the
 physical sector is demonstrated.
 \end{abstract}
\section{Introduction}\par
 \par The issue of dynamical systems with constraints dominates in the modern physics. Gauge theories describing all the known
  interactions (gravitational, electro-weak, strong)
 are typical examples of such systems.
   As it is well known \cite{dirac}, there are two
   classes of constraints: the first and the second ones.
 Appearance of  constraints  results in reduction of the phase space of the initial
  theory but  mechanisms of the reduction differ
  for constraints from  different classes.
   To elucidate the problem
  we can  take the simplest case and consider the first class
  constraints as conditions on some canonical momenta
    $p_{r+1}=\ldots=p_{r+s}=0$ ($s$ first class constraints, $r+s=n$), and second class constraints  ---  as conditions on  momenta and the canonically conjugate coordinates
  $p_{r+1}=\ldots=p_{r+s}=0$, $q_{r+1}=\ldots=q_{r+s}=0$ ($2s$ second class constraints).
  In both cases dimension of the physical phase space (PS) becomes
  $2(n-s)$, because  in the case of the
  first class constraints  the variables $q_{r+1},\ldots$,$q_{r+s}$ are also non-physical (they remain arbitrary). So,
  if we consider theory with $n$ degrees of freedom, then in both cases the  dimension of the physical
   phase space is equal to $2n-2s$. In the classical theory this  does not cause problems
   because for the systems with second class constraints there is
   no  arbitrariness while for the systems with first class
   constraints the arbitrariness can be removed by  gauge fixing
   \begin{equation} q_{r+1}=\ldots=q_{r+s}=0.\label{1}\end{equation}
   It is admissible, because the theory  does not impose any
   restrictions on evolution of the variables (\ref{1}). Problems occur
  after transition to the quantum description. One cannot
  require the operator equations   $\hat{p}_{r+1}\ldots=\hat{p}_{r+s}=0$
    to be fulfilled  in the case of the first class constraints
   because  it contradicts to commutation relations
   $[\hat{q}_{r+k},\hat{p}_{r+l}]=i\hbar\delta_{kl},1\leq k,l\leq s.$
   And of course, one cannot require it in presence of additional
   conditions
  $\hat{q}_{r+1}=\ldots=\hat{q}_{r+s}=0$ (the second class constraints).
  In the former case there is natural way out \cite{dirac}: it is sufficient to
  require realization of
constraints on physical  vectors $\Psi_{ph}$
\begin{equation}
  \hat{p}_{r+1}\Psi_{ph}=...=\hat{p}_{r+s}\Psi_{ph}=0.\label{2}\end{equation}
  But this recipe does not suite for the second class constraints: conditions
    $\hat{p}\Psi_{ph}=\hat{q}\Psi_{ph}=0$ are incompatible with
    canonical commutation relations  $[\hat{q},\hat{p}]=i\hbar$.  Several ways out there was
    proposed. Two of them are based  on the fact, that variables  $q_{r+k},p_{r+k},
   1\leq k\leq s$, are non-physical. So, it is supposed that it is possible  to change  their
  dynamics arbitrary.  Dirac \cite{dirac} proposed to change the Poisson brackets
   $\{\Phi_{i},\Phi_{j}\}$;
    for the second class constraints (i.e. $\Phi_{i}=0$
  $det\{\Phi_{i},\Phi_{j}\}\neq0$) he proposed to use new brackets
  \begin{equation}\{f,g\}\rightarrow\{f,g\}_{D}=\{f,g\}-\{f,\Phi_{i}\}(\{\Phi_{i},\Phi_{j}\})^{-1}\{\Phi_{j},g\}. \label{3}\end{equation}
   The Dirac brackets
   $\{f,g\}_{D}$ are equal to zero for constraints $\Phi_{i}$, i.e.
   the latter
   become
   the first class constraints and it is possible to use conditions
    (\ref{2}) $\hat{\Phi_{i}}\Psi_{Ph}=0$. Thus, non-physical canonical conjugate variables $q,p$
    become
    independent (non-physical) variables. Variables canonically
    conjugated them are ignored \cite[p. 130]{psh}.\par
    In the recipe [3--5] one doubles the number of non-physical variables;
      it is postulated that the Poisson brackets of  all non-physical
     canonical  variables of the original system
     are equal to zero (" abelian conversion"), while  new variables are  assumed be canonically
     conjugated to them.
   Then, it is again possible to use recipe  (\ref{2}) for constraints  $\Phi_{i}$.
    In fact, the Dirac recipe is accomplished by introducing  canonically conjugated
    partners for non-physical "momenta" \ \ $\Phi_{i}$.\par
    Restriction of motion in the configuration space (for example, on some
    hypersurface in Euclidean space defined by conditions $\varphi_{i}(q_{1},\ldots,q_{n})=0 ,  i=1,\ldots,s<n )$
    is the typical reason for the second class constraints occurrence.
      In this approach quantum mechanics (QM) on a hypersurface can be considered as a
      limiting case of $n$---dimensional QM in an infinitesimally thin layer
      surrounding the hypersurface. Such a method of quantization in curved spaces is called "the thin layer
    method". It appears in two forms. In the first one it is required that the wave functions
     become zero on borders of the layer \cite{jk}. In this case decreasing  of thickness of the layer
     to zero leads to occurrence of states with infinite energy, and "renormalization" \, of energy
     is needed for transition to  QM on the hypersurface.
      In the second method \cite{dacosta} one introduces an oscillator potential in directions normal to
     the  hypersurface; when the elasticity coefficient tends to infinity wave function turns out confined
       on the hypersurface.
    It was found in \cite{dacosta}
    that as a result some function  $V_{q}$ ("quantum potential")  is added to the Beltrami---Laplace operator
     on the hypersurface (i.e. to kinetic energy operator).
      It turns out (and this is extremely important) that the potential
      $V_{q}$ depends both on intrinsic and extrinsic curvatures
      of the hypersurface.
     By itself this fact is quite satisfactory, because wave function is a non-local object.
     It is remarkable  that  extrinsic curvature also influences motion of a quantum particle.
     In the classical mechanics  motion of a point particle
     depends
     only on intrinsic geometry.\par
     Path integral method gives quantum potential
    $V_{q}=\frac{\hbar^{2}R}{12}$, where $R$ --- scalar curvature
    of space \cite{Witt}. And only mechanics with the second class
    constraints allows to reveal the  important fact that quantum potential depends
     not only on intrinsic but also on \emph{extrinsic} curvature.
    In section 2 the results  of different methods are listed for the simplest case
    --- a particle on a sphere. All the mentioned methods give
     different results; it means that at least two of them are incorrect.
     In sec. 3 a new method of quantization, naturally following
      from rules of quantization of systems with the first class constraints is presented.
       In sections 4,  5 it is shown, that for a plane curve   and surfaces in
       3-dimensional Euclidean space the  method gives the same results
       as the thin layer method. In sec. 6 the influence of the non-physical sector
        of a system on the physical one is discussed. It is shown, that quantum potential depends
        on space, in which e.g. a  curve is taken: a circle on a plane or on
        a sphere. Thus, in QM the non-physical sector
         influences  the physical one.
    \section{Particle on a sphere --- three recipes of quantization}
In this section we present results of quantization by three
various methods for simplest case --- a particle on the sphere of
radius $R$ in  $R^{n}$ \cite{ksh}.

1. \emph{The Dirac method}.
Constraints:$$\Phi_{1}=\vec{x}^{2}-R^{2},$$
$$\Phi_{2}=(\vec{p},\vec{x}).$$
 Hamiltonian:
 $$\hat{H}=-\frac{\hbar^{2}}{2}\triangle_{n-1}+\frac{\hbar^{2}n^{2}}{8R^{2}}, $$
  where $\triangle_{n-1}$ --- the Beltramy---Laplace operator on
  the sphere. Quantum potential is:
  \begin{equation}V_{q}^{D}=+\frac{\hbar^{2}n^{2}}{8R^{2}}.\end{equation}

  2. \emph{The Abelian conversion method}. Constraints $\Phi_{1}$, $\Phi_{2}$
  are postulated abelian;  new auxiliary variables $Q$ and $P$, canonically
   conjugated to them  are introduced.
$$\{Q,\Phi_{1}\}=\{P,\Phi_{2}\}=1,
 \{Q,\Phi_{2}\}=\{P,\Phi_{1}\}=0.$$

    Then new constraints
  $$\sigma_{1}=\Phi_{1}+P,$$
  $$\sigma_{2}=\Phi_{2}+2\vec{x}^{2}Q,$$ are in involution:
  $$\{\sigma_{1},\sigma_{2}\}=0.$$
 The Hamiltonian is
  $$H=\frac{1}{2(\sigma_{1}^{2}+R^{2})}(\sigma_{2}^{2}+L_{a}^{2}),$$
  where $L_{a}=x_{i}p_{j}-x_{j}p_{i}$$,\ \ a\equiv(ij)$ --- components
  of the angular momentum operator. Quantum potential is zero:
  $$V_{q}^{AC}=0.$$

  3. \emph{The thin layer method}. Motion on a surface is considered  as a  motion in
  the Euclidean space between two parallel (equidistant) surfaces
  when distance between the surfaces tends to zero \cite{jk}. The energy of the
  system tends in this case to infinity. More attractive looks the idea of
taking "squeezing" \-  potentials
  (e.g. by introducing an oscillator potential
$V=\frac{1}{2}\gamma\vec{x}_\bot^{2},\gamma\rightarrow\infty$, in
normal directions) \cite{dacosta}. Then for a particle on a
surface in 3-dimensional Euclidean space one obtains quantum \,
potential

$$V_{q}^{ThL}=-\frac{\hbar^{2}}{2}(H^{2}-K)=-\frac{\hbar^{2}}{8}(\frac{1}{R_{1}}-\frac{1}{R_{2}})^{2},$$
   where $H$ --- the mean curvature, $K$ --- the Gaussian curvature,
   $R_{1}^{-1},R_{2}^{-1}$ --- the principal  curvatures of the surface. For a motion on a circle
   ($R_{1}=R,R_{2}=\infty$) we have $$V_{q}^{ThL}=-\frac{\hbar^{2}}{8R^{2}}.$$
   For a motion on a sphere ($R_{1}=R_{2}$) we have $V_{q}^{ThL}=0$.
\section{Reduction to the case of first class constraints }
  The following method naturally follows from the first class constraints quantization
  rules.
   Let $q_{i}, i=1,\ldots,n$, be curvilinear orthogonal coordinates in
   $R^{n}$ so that conditions $q_{r+k}-C_{k}=0$, $C_{k}=const$,
      $k=1,\ldots,s$, $r=n-s$, define a hypersurface with dimension
      $n-s$, i.e. coordinates $q_{r+1},\ldots,q_{n}$ are normal to
      the hypersurface,
   and $p_{r+1},\ldots,p_{n}$ are momenta canonically conjugated to them. Quantization can be done in two steps.

   I. On solutions of the Schr\"{o}dinger equation \begin{equation}-\frac{\hbar^{2}}{2}\triangle
\Psi(q_{1},\ldots,q_{n})=E\Psi(q_{1},\ldots,q_{n}), \label{free0}
\end{equation} we impose conditions \begin{equation}\hat{P}_{r+k}\Psi(q_{1},\ldots,q_{n})=0,\ \ k=1,\ldots,n-s, \label{condition}\end{equation}
where $\triangle$ --- the Laplace operator in $R^{n}$ in
curvilinear coordinates, and the momentum operators are:
\begin{equation}
\hat{P_{j}}=\frac{\hbar}{i}g^{-\frac{1}{4}}\hat{\partial
_{j}}g^{\frac{1}{4}},\ \ \hat{P}_{j}^{+}=\hat{P}_{j},\ \
j=1,\ldots,n,\label{imp}\end{equation}  $\hat{P_{j}}$ are
Hermitean in the Hilbert space with the inner product
$(f_{1},f_{2})=\int d^{n}q\sqrt{g}\bar{f_{1}}(q)f_{2}(q)$ where
$g$ is determinant of the metric tensor. After imposing conditions
(\ref {condition}) the wave functions actually do not depend on
non-physical coordinates
$$\Psi_{ph}(q_{1},\ldots,q_{n})=g^{-\frac{1}{4}}\Phi(q_{1},\ldots,q_{r}),$$ and it is
possible to consider the latter not as dynamical variables but as
parameters, because now in the normalization condition one
integrates only on physical variables
\begin{equation}\label{zac1}
    \int d^{r}q\sqrt{g}\mid \Psi_{ph} \mid ^{2}=1.
\end{equation}
II. Further, it is necessary: 1) substitute in Jacobians $ \sqrt
{g} $ all $ q _ {r+ k}, \ \ k = 1,2, \ldots, s $ by $ C _ {k} $
(resolve constraints  $ q _ {r + k} -C _ {k} = 0 $); \ \ 2) move
in the Hamiltonian all operators of non-physical momenta $ \hat{P}
_ {r + k} $  to the right (taking into  account the commutation
relationes) and  put them equal to zero (see (\ref{condition})); \
\ 3) replace in the final operator   all $ q _ {r + k} $ by $ C _
{k} $. It turns out, that this method gives the same results as
 the thin layer method. The  method looks natural, because in the
case of the first class constraints sometimes "additional
conditions" \,(such as conditions (\ref {1})), are introduced
("gauge fixing").

In the following sections we shall see, that this recipe in the
natural way gives  the Sсhr\"{o}dinger equation on hypersurface
with quantum potential depending both on intrinsic (scalar
curvature) and extrinsic geometry (curvature of plane curve in
$R^{2}$ and mean curvature $H$ of a surface in $R^{3}$). This
quantum potential depends not only on geometry
 of hypersurface, but also on geometry of imbedding space as shows an example of a circle in
$R^{2}$ and on a sphere $S^{2}$ (section 6). Notice, that we do
not introduce squeezing potentials, as in \cite{jk,dacosta}, which
should keep a particle on the hypersurface. The idea of this
"two-step reduction"\, recipe was introduced in \cite{LV}. The
recipe (\ref{free0}),\,(\ref{condition}) was considered
\cite{kleinert}, but was rejected in favor of the abelian
conversion method.

\section{Quantum potential on a plane curve.}
We shall illustrate the recipe (\ref {free0}), (\ref {condition})
first for a plane curve with
 curvature $k (s)$, where $ s $ is length of an arc. We use special
 curvilinear coordinates (" coordinates of a thin layer ")
 $ q _ {1}, q _ {2} $, where $ q _ {1} $ --- length of the arc of a curve from some
"zero"\, point, $ q _ {2} $ --- distance between  the curve and a
plane point. The first quadratic
  form in such (semi-geodesic) coordinates becomes
 $$ dl ^ {2} = dq _ {2} ^ {2} + gdq _ {1} ^ {2}, $$
 where $$ g = [1 - q _ {2} k (q _ {1})] ^ {2}. $$
  The lines $ q _ {2} = const $  are   equidistant  curves parallel to the
  original curve.\ \ Lines $ q _ {1} = const $ are normal to it.
  The Sсhr\"{o}dinger equation for a free particle on a
  plane reads
\begin{equation} -\frac{\hbar^{2}}{2}\triangle
\Psi(q_{1},q_{2})=E\Psi(q_{1},q_{2}), \label{free}\end{equation}
and  the condition on the wave function is \begin{equation}
\hat{P_{2}}\Psi(q_{1},q_{2})=0, \label{condition0}\end{equation}
where the momentum operator  $ \hat {P} _ {2} $ is given by
equation (\ref {imp}).
\par It follows from (\ref{imp}), (\ref {condition0})  that the wave function
has the form:\begin {equation} \Psi (q _ {1}, q _ {2}) = g ^
{-\frac {1} {4}} \Phi (q _ {1}) .\end {equation}\par Substituting
this expression in (\ref {free}) we obtain \begin {equation}
-\frac{\hbar^{2}}{2}\left(g ^ {-\frac {1} {2}}
\partial _ {1} [g ^ {\frac {1} {2}} g ^ {-1} \partial _ {1} (g ^
{-\frac {1} {4}} \Phi)] - (g ^ {-\frac {1} {4}} \Phi) [g ^ {-\frac
{1} {4}} \partial _ {2} ^ {2} g ^ {\frac {1} {4}}]\right) = E (g ^
{-\frac {1} {4}} \Phi),
\end {equation}
$$ g ^ {-\frac {1} {4}} \partial _ {2} ^ {2} g ^ {\frac {1} {4}} = -\frac {1} {4} \frac {k ^ {2}} {(1 - q _ {2} k) ^ {2}}, $$
 Taking $ q _ {2} = 0, $ we reproduce the result of papers
 \cite {jk, dacosta}: $$
-\frac {\hbar^{2}} {2} \partial _ {1} ^ {2} \Phi-\frac {\hbar^{2}}
{8} k ^ {2} \Phi = E\Phi, $$
 i.e. quantum potential is equal to
\begin {equation} V _ {q} = -\frac {\hbar ^ {2}} {8} k ^ {2} .\end {equation}
\section {\textbf{Quantum potential for a particle on a surface in three-dimensional
space}}
\par Let's consider a surface $ \Gamma $ in space $ R ^
{3} $. We introduce coordinates $ q _ {1}, q _ {2} $ such that the
first and the second quadratic forms become diagonal
$$ ds ^ {2} = h _ {1} ^ {2} dq _ {1} ^ {2} + h _ {2} ^ {2} dq _ {2} ^ {2}. $$
 Following authors  \cite {jk, dacosta}, we introduce in the thin
 layer
 special  curvilinear orthogonal
coordinates (coordinates of a thin layer). Equation $ \vec {r} =
\vec {r} (q _ {1}, q _ {2}) $ parametrizes   the
 surface. A point in $ R ^ {3} $ is characterized by coordinates
$ (q _ {1}, q _ {2}) $ of the surface $ \Gamma $ and by the
distance $ q _ {3} $ from it: $$\vec {R} (q _ {1}, q _ {2}, q _
{3}) = \vec {r} (q _ {1}, q _ {2}) + q _ {3} \vec {n} (q _ {1}, q
_ {2}),
$$ where $ \vec {n} $ is a normal  to the surface. Then the
metrics in $ R ^ {3} $ is: $$ ds ^ {2} = H _ {1} ^ {2} dq _ {1} ^
{2} + H _ {2} ^ {2} dq _ {2} ^ {2} + H _ {3} ^ {2} dq _ {3} ^ {2},
$$ where $$ H _ {1} = h _ {1} (1 - q _ {3} k _ {1}), \ \ H _ {2} =
h _ {2} (1 - q _ {3} k _ {2}), \ \ H _ {3} = 1$$  and
$k_{1},k_{2}$ are principal curvatures.
\par The metrics of
the surfaces $ q _ {3} = const $, parallel to the surface $ \Gamma
$, is defined by  $$ ds ^ {2} = H _ {1} ^ {2} dq _ {1} ^ {2} + H _
{2} ^ {2} dq ^ {2} _ {2}. $$\par The stationary Sсhr\"{o}dinger
equation  for a free particle in $ R ^ {3} $ reads
$$-\frac {\hbar^{2}} {2} \triangle _ {3} \Psi (q _ {1}, q _ {2}, q _ {3}) = E\Psi (q _ {1}, q _ {2}, q _ {3}). \label {free2} $$
 We  demand $$\hat {P _ {3}} \Psi = 0, $$ where
$ \hat {P _ {3}} $
$$\hat {P _ {3}} = \frac {\hbar} {i} G ^ {-\frac {1} {4}} \hat {\partial _ {3}} G ^ {\frac {1} {4}}, $$
$$ G ^ {\frac {1} {2}} = H _ {1} H _ {2} H _ {3} = h_{1}h_{2}(1 + q _ {3} (k _ {1} + k _ {2}) + k _ {1} k _ {2} q _ {3} ^ {2}) = h_{1}h_{2}(1 + 2q _ {3} H + Kq _ {3} ^ {2}), $$
where $K$ and $H$ are the  Gauss and mean curvatures of the
surface $\Gamma$. The wave function $\Psi$ and the Laplace
operator become
$$\Psi(q_{1},q_{2},q_{3})=G^{-\frac{1}{4}}\Phi(q_{1},q_{2}),\label{factor}$$
$$\triangle_{3}=\triangle_{2}+G^{-\frac{1}{2}}\hat{\partial_{3}}(G^{\frac{1}{2}}\hat{\partial_{3}}),$$
 where $\triangle_{2}$ --- the Beltrami---Laplace operator on the surfaces $q_{3}=const$, i.e.
$$\triangle_{3}\Psi=\triangle_{2}\Psi+\Psi G^{-\frac{1}{4}}\partial_{3}(G^{\frac{1}{2}}\partial_{3}G^{-\frac{1}{4}}).$$

Taking $q_{3}=0$ we reproduce the result of papers
\cite{jk,dacosta} :$$
-\frac{\hbar^{2}}{2}\triangle_{2}\Psi(q_{1},q_{2})-\frac{\hbar^{2}}{2}(H^{2}-K)\Psi(q_{1},q_{2})=E\Psi(q_{1},q_{2})\label{main}
,$$ i.e. quantum potential is equal to
:\begin{equation}V_{q}=-\frac{\hbar^{2}}{2}(H^{2}-K)\label{curve}.\end{equation}
\par
\section{Quantum potential and geometry of embedding space}
 As it was noticed in Introduction  quantum potential
depends on geometry of imbedding space. To demonstrate this we
consider a circle  on a sphere and on a plane.

On the sphere of radius $R$ take  a circle $\theta=const$
($\theta$,$\phi$ are spherical coordinates). Here $\phi$  is the
physical coordinate, while  $\theta$ --- the non-physical one.
Metrics is given by $ds^{2}=R^{2}(d\theta^{2}+\sin^{2}\theta
d\phi^{2})$, and $g=R^{4}\sin^{2}\theta$. According to Eq.
(\ref{condition}) physical states are:
$\Psi(\phi,\theta)=g^{-\frac{1}{4}}\Phi(\phi)=R^{-1}(\sin\theta)^{-\frac{1}{2}}
\Phi(\phi)$. The Schr\"{o}dinger equation (\ref{free0}) takes
form:
$$-\frac{\hbar^{2}}{2}\left(R^{-2}\sin^{-2}\theta\frac{\partial^{2}\Psi}{\partial\theta^{2}}+\Psi R^{-2}\sin^{-\frac{1}{2}}\theta \frac{\partial}{\partial\theta}[\sin\theta\frac{\partial}{\partial\theta}\sin^{-\frac{1}{2}}\theta]\right)=E\Psi$$
and quantum potential is:
\begin{equation}\label{66}V_{q}=(-\frac{\hbar^{2}}{2})R^{-2}\sin^{-\frac{1}{2}}\theta
\frac{\partial}{\partial\theta}[\sin\theta\frac{\partial}{\partial\theta}\sin^{-\frac{1}{2}}\theta]=-\frac{\hbar^{2}}{8R^{2}}[1+\frac{1}{\sin^{2}\theta}].\end{equation}
 On the other hand,   in the case of the circle with
$R_{c}=R\sin\theta$ on a plane we had
\begin{equation}\label{67}V_{q}=-\frac{\hbar^{2}}{8(R\sin\theta)^{2}}=-\frac{\hbar^{2}}{8R_{c}^{2}}.\end{equation}
  We see  that
quantum potential depends on the  curvature of the imbedding
space. Of course  in the limit $R\rightarrow
\infty,\theta\rightarrow 0,R\sin \theta\rightarrow R_{c}$ the
results (\ref{66}) and (\ref{67}) coincide. In the framework of QM
the difference between  Eqs.\,(\ref{66}) and (\ref{67}) is both
natural and desirable: wave function is a non-local object and we
began with formulation of the problem in the imbedding space.

\section{Conclusion}
The analysis of existing quantization methods  of systems with
second class constraints shows that this problem is  less trivial
than similar problem in the case of first class constraints. These
 methods of quantization can be divided into two groups.

To the first group we attribute the methods   admitting that the
non-physical sector cannot influence  the physical one, so one can
arbitrary  change dynamics of the non-physical variables: for
example to replace their Poisson brackets by Dirac brackets
\cite{dirac} or to set Poisson brackets equal to zero for all
non-physical canonical variables \cite {faddeev,batalin}, adding
extra non-physical variables. Application of these recipes to the
elementary systems gives the different results that means: in the
quantum theory one cannot change arbitrary the dynamics of the
non-physical sector.

The methods in the second group  do not modify dynamics of the
non-physical sector.  So, in papers \cite {jk, dacosta} the
quantization is made in space of all variables, both physical and
non-physical ones ("the thin layer method"). In the  limit of
vanishing
 thickness of the "layer" \, one gets  quantum theory on the physical subspace. The final result
differs from results, received by methods from the first group.
Unfortunately this recipe requires large auxiliary work. It turns
out however that there is a more direct way
--- method of "reduction the problem to the first class constraints problem"\, resulted in sec. 3.

Let's address in conclusion  the question of influence of
non-physical sector on the physical one. We see, that  the
non-physical sector influences the physical one in all the methods
of quantization. It means  that the problem of  quantization  on,
say, a curve is by itself set incorrectly. It is necessary to
specify space, in which the curve is imbedded. For example, in the
case of a plane, the quantum potential $V_{q}$ is given by  (\ref
{67}) (radius of a circle is equal to $ R\sin\theta $), and if the
circle is on the sphere of radius $ R $  the potential $V_{q}$ is
given by (\ref {66}). The result looks paradoxical only from the
point of view of the classical theory. In quantum theory the
motion of particles is described by wave functions, and it is not
surprising that the motion on a circle depends on the outer space
--- sphere or plane. In the first case the wave function does not
depend on one of the spherical coordinates (angle $ \theta $),
while in second one
--- on the radial variable.

Conclusion: in  quantum mechanics  description of motion  in
curved spaces by itself, i.e. ignoring the imbedding space, is
senseless.

\end{document}